\documentclass{article}

\usepackage{arxiv}

\usepackage[utf8]{inputenc} 
\usepackage[T1]{fontenc}    
\usepackage{hyperref}       
\usepackage{url}            
\usepackage{booktabs}       
\usepackage{amsfonts}       
\usepackage{nicefrac}       
\usepackage{microtype}      
\usepackage{lipsum}
\usepackage{graphicx}

\usepackage{cite}
\usepackage{amsmath,amssymb,amsfonts}%
\usepackage{amsthm}%
\usepackage{mathrsfs}%
\usepackage{textcomp}
\usepackage{algpseudocode}%
\usepackage{algorithm}%
\usepackage{algorithmicx}%
\usepackage{listings}%
\usepackage{stfloats}
\usepackage{lineno,hyperref}

\algnewcommand\algorithmicforeach{\textbf{for each}}
\algdef{S}[FOR]{ForEach}[1]{\algorithmicforeach\ #1\ \algorithmicdo}

\newboolean{showcomments}
\setboolean{showcomments}{true}
\ifthenelse{\boolean{showcomments}}
{\newcommand{\nb}[2]{
\fbox{\bfseries\sffamily\scriptsize#1}
{\sf\small$\blacktriangleright$\textit{#2}$\blacktriangleleft$}
}
}
{\newcommand{\nb}[2]{}
}

\graphicspath{ {./images/} }

\title{Adaptive Temporal Fusion Transformers for Cryptocurrency Price Prediction}

\author{
 Arash Peik \\
  Department of Computer Engineering\\
  Yazd University\\
  Yazd, Iran \\
  \texttt{peik@stu.yazd.ac.ir} \\
   \And
 Mohammad Ali Zare Chahooki \\
  Department of Computer Engineering\\
  Yazd University\\
  Yazd, Iran \\
  \texttt{chahooki@yazd.ac.ir} \\
  \And
 Amin Milani Fard \\
  Department of Computer Science\\
  New York Institute of Technology\\
  Vancouver, Canada \\
  \texttt{amilanif@nyit.edu} \\
  \And
 Mehdi A. Sarram \\
  School of Coumputing and Information\\
  Yazd University\\
  Yazd, Iran \\
  \texttt{mehdi.sarram@yazd.ac.ir} \\
}

\begin{document}
\maketitle
\begin{abstract}
Precise short-term price prediction in the highly volatile cryptocurrency market is critical for informed trading strategies. Although Temporal Fusion Transformers (TFTs) have shown potential, their direct use often struggles in the face of the market’s non-stationary nature and extreme volatility. This paper introduces an adaptive TFT modeling approach leveraging dynamic subseries lengths and pattern-based categorization to enhance short-term forecasting. We propose a novel segmentation method where subseries end at relative maxima, identified when the price increase from the preceding minimum surpasses a threshold, thus capturing significant upward movements, which act as key markers for the end of a growth phase, while potentially filtering the noise. Crucially, the fixed-length pattern ending each subseries determines the category assigned to the subsequent variable-length subseries, grouping typical market responses that follow similar preceding conditions. A distinct TFT model trained for each category is specialized in predicting the evolution of these subsequent subseries based on their initial steps after the preceding peak. Experimental results on ETH-USDT 10-minute data over a two-month test period demonstrate that our adaptive approach significantly outperforms baseline fixed-length TFT and LSTM models in prediction accuracy and simulated trading profitability. Our combination of adaptive segmentation and pattern-conditioned forecasting enables more robust and responsive cryptocurrency price prediction.
\end{abstract}


\section{Introduction}
\label{sec:introduction}

The cryptocurrency market, characterized by its high volatility, complex dynamics, and non-stationary nature, presents both significant opportunities and substantial risks for traders and investors \cite{corbet2019cryptocurrencies}. Accurate forecasting of price movements, particularly over short time horizons, is crucial to developing profitable trading strategies and managing risk \cite{angela2020factors,sovbetov2018factors}. However, predicting these fluctuations is a formidable task due to the inherent complexities and sensitivity of the market to numerous factors \cite{angela2020factors}.

In recent years, machine learning and deep learning models have emerged as powerful tools to tackle financial time series forecasting \cite{yamin2023cryptocurrency}. Various studies have examined the performance of different algorithms; for instance, Akyildirim et al. \cite{akyildirim2023forecasting} assessed machine learning models for Bitcoin mid-price movement, while Chen et al. \cite{chen2019predicting} applied classic machine learning to Ethereum price prediction. Deep learning models, especially Recurrent Neural Networks (RNNs) such as Long Short-Term Memory (LSTM), have been widely adopted due to their ability to capture temporal dependencies \cite{fazlollahi2023predicting, song2020time}. Fazlollahi and Ebrahimijam \cite{fazlollahi2023predicting} utilized a multi-input LSTM with technical indicators, and Bouteska et al. \cite{bouteska2024cryptocurrency} provided comparative analysis of ensemble and deep learning methods. More recently, transformer-based architectures, originally developed for natural language processing \cite{vaswani2017attention}, have demonstrated strong capabilities. The Temporal Fusion Transformer (TFT) \cite{lim2021temporal}, in particular, is notable for its ability to model complex long-range dependencies and incorporate various input types, achieving state-of-the-art results in several forecasting benchmarks \cite{hu2021stock}.

Despite the power of models such as TFT, their direct application to raw cryptocurrency price data, often using fixed-length sliding windows, can be suboptimal. While computationally convenient, this approach may arbitrarily truncate meaningful market phases or do not adapt to the varying durations and characteristics of different market patterns and regimes \cite{peik2024leveraging,Peik2025Enhancing}. Training a single model on the entire series might also prevent specialization in learning distinct patterns occurring during different market conditions. Recognizing this limitation, in our previous work \cite{Peik2025Enhancing} we explored enhancing TFT performance by categorizing fixed-length subseries based on their internal patterns and training specialized models for each category. Although this showed improvements over the baseline TFT and LSTM models, the rigidity of fixed-length segmentation remained a constraint.

Other research directions have also explored partitioning time series or integrating diverse data. For instance, regime-switching models aim to identify distinct market states or regimes. RHINE \cite{xu2024rhine} employs kernel representation learning to discover and forecast regimes based on inter-series dependencies. This would leverage the specific strengths of TFTs in modeling intricate temporal dynamics directly within each identified behavioral category. Another active area involves multimodal fusion, combining price data with external information such as technical indicators or sentiment analysis from news and social media, as seen in models such as MFB \cite{HAN2025125515} and others \cite{farimani2021leveraging,farimani2022investigating,farimani2024adaptive,Farimani2025Financial,chen2024deep}. Although  valuable, these approaches often focus on broader regime identification or data fusion rather than adapting the core analysis window to the intrinsic structure of price movements for pattern-conditioned forecasting.

\textbf{Contributions.} This article presents a substantial extension over our recent work \cite{Peik2025Enhancing}. Our preliminary research demonstrated that while categorizing \textit{fixed-length} time series windows improved prediction accuracy, it was constrained by the rigidity of such arbitrary segmentation. To overcome this limitation, in this work we introduce \textit{Adaptive TFT Modeling}, a novel approach designed to dynamically align the analysis with the natural rhythm of the market.

Our methodology is built on two core innovations:

\begin{enumerate}
    \item We replace fixed windows with an adaptive segmentation algorithm that partitions the volatility time series into \textit{variable-length} subseries. The boundaries of these segments are not predetermined but are demarcated by significant relative maxima, identified when a price increase from the preceding minimum surpasses a specific threshold. 
    
    \item We introduce a pattern-conditioned forecasting strategy in which the predictive model for a given subseries is selected based on the characteristic pattern observed at the end of the preceding subseries. 

\end{enumerate}

Together, these techniques allow our models to learn from more coherent market phases and make context-aware predictions, leading to a more robust and responsive forecasting framework.

In our pattern-conditioned forecasting strategy, instead of categorizing a subseries based on its own structure, we categorize it based on the fixed-length pattern observed at the \textit{end} of the \textit{preceding} subseries. This implies that the category represents typical market behavior expected to follow a specific concluding pattern. We then train a specialized TFT model for each category, leveraging the architecture's power to predict the evolution of the current, unfolding subseries, conditioned on the pattern that concluded the previous market phase. This allows TFT models to become experts in predicting the consequences of specific, recent market dynamics.

We hypothesize that this combination of adaptive, market-driven segmentation and pattern-conditioned specialized forecasting will lead to more accurate and responsive short-term predictions compared to fixed-length methods or standard TFT models.

A foundational principle of our approach, carried over from our initial work \cite{Peik2025Enhancing}, is the specialization of predictive models based on recurring behavioral patterns rather than simple geometric similarity. Traditional clustering methods, such as $k$-means, often group subseries that are close in the $n$-dimensional space but exhibit fundamentally different market dynamics. Our pattern-based categorization, in contrast, ensures that each category represents a distinct and interpretable market behavior.

In addition, this work evolves beyond our architecture proposed in \cite{peik2024leveraging,Peik2025Enhancing} by eliminating the need for a separate selector model (such as a Markov chain or LSTM) to predict the next market phase. In our new adaptive framework, the prediction process is more self-contained: the specialized TFT model, selected based on the concluded pattern of the last subseries, is solely responsible for forecasting the evolution of the current unfolding market phase. This design streamlines the prediction pipeline and better leverages the TFT's inherent capability to model complex conditional dependencies.

Overall, our work makes the following key contributions:

\begin{enumerate}
    \item A novel adaptive segmentation algorithm for cryptocurrency time series based on thresholded relative maxima.
    \item A pattern-conditioned categorization framework where the category of a subseries is determined by the end-pattern of the preceding subseries.
    \item An adaptive forecasting methodology employing specialized TFT models trained for each category to predict variable-length subseries evolution.
    \item Empirical validation demonstrating the superior performance of the proposed method in accuracy and simulated trading profitability compared to baseline models on ETH-USDT data.
\end{enumerate}

The remainder of this paper is organized as follows: Section~\ref{sec:related_work} discusses related work in more detail. Section~\ref{sec:proposed_approach} elaborates on our proposed methodology. Section~\ref{sec:experiments} describes the experimental setup and presents the results. Section~\ref{sec:discussion} discusses the findings and limitations, and Section~\ref{sec:conclusion} concludes the paper and outlines future research directions.

\section{Related Work}
\label{sec:related_work}

The prediction of financial time series, particularly in the volatile cryptocurrency market, has been a subject of extensive research, employing a wide range of methodologies from classical statistics to advanced deep learning. This section reviews the relevant literature on time series forecasting models, segmentation and regime analysis, pattern recognition, and multimodal approaches, contextualizing our proposed adaptive methodology.

\subsection{Time Series Forecasting Models}

Traditional machine learning algorithms, such as Support Vector Machines (SVM), Random Forests, and K-Nearest Neighbors (KNN), have been applied to stock and cryptocurrency market prediction \cite{kumar2018comparative, akyildirim2023forecasting, chen2019predicting}. However, their ability to capture complex temporal dependencies is often limited.

Deep learning models, particularly Recurrent Neural Networks (RNNs) and their variants such as Long Short-Term Memory (LSTM) and Gated Recurrent Units (GRU), marked a significant advancement \cite{song2020time, fazlollahi2023predicting}. LSTMs, with their gating mechanisms, can mitigate the vanishing gradient problem and learn long-range dependencies, making them suitable for sequential data such as price series. Hybrid models that combine Convolutional Neural Networks (CNNs) for feature extraction and RNNs for sequence modeling have also been explored \cite{kang2022cryptocurrency}.

More recently, transformer-based architectures \cite{vaswani2017attention}, leveraging self-attention mechanisms, have revolutionized sequence modeling. Models such as the Temporal Fusion Transformer (TFT) \cite{lim2021temporal} adapt the transformer architecture specifically for time series forecasting. TFT incorporates gating mechanisms (GRNs), variable selection networks, and interpretable multi-head attention to handle static covariates, known future inputs, and observed past inputs simultaneously, achieving state-of-the-art performance on various benchmarks \cite{hu2021stock}. Other Transformer variants such as Informer \cite{zhou2021informer} focus on efficiency for very long sequence forecasting. Our work utilizes TFT as the core predictive engine due to its proven capabilities in complex time series modeling, achieving state-of-the-art results in several forecasting benchmarks \cite{electronics12224656}.

\subsection{Time Series Segmentation and Regime Analysis}

Standard time series analyses often employ a fixed-length sliding window approach. While simple, they can be suboptimal as market dynamics rarely conforms to fixed durations. Recognizing this, various approaches have sought to partition time series more meaningfully.

Our previous work \cite{Peik2025Enhancing} explored categorizing \textit{fixed-length} subseries based on their internal volatility patterns (e.g., sequences of up/down movements) and training specialized TFT models for each category. This demonstrated the benefit of specialization; however, was still constrained by the fixed segment length.

Another major line of research involves \textit{regime-switching models}. These models assume that the underlying data generation process switches between a finite number of states or "regimes," each with different statistical properties. Classical approaches often use Markov Switching Models (MSMs) \cite{hamilton1989new}, where transitions between regimes are governed by a Markov chain. Extensions include models incorporating switching conditional variance (e.g., Markov Switching GARCH \cite{cai1994markov, ardia2019markov}) or time-varying transition probabilities \cite{bazzi2017time}.
However, many MSM approaches require the number of regimes to be specified a priori and often model individual series in isolation.

More recent works attempt to discover regimes dynamically or across multiple series. Matsubara and Sakurai proposed methods for forecasting patterns or transitions between known regimes in co-evolving streams \cite{matsubara2019dynamic}. Xu et al. recently introduced RHINE \cite{xu2024rhine}, which uses kernel representation learning and self-representation with a block diagonal regularizer to automatically discover regimes and model nonlinear interdependencies across multiple time series without prior regime labels. 
While RHINE focuses on regime discovery and forecasting regime switches based on inter-series dependencies, our work differs by focusing on adaptive segmentation based on intra-series turning points ($T_h$ condition) and using the end-pattern of the \textit{preceding} segment to condition the prediction of the \textit{subsequent} segment's evolution via specialized TFTs.

Our proposed adaptive segmentation based on thresholded relative maxima (Section~\ref{subsec:adaptive_segmentation}) offers an alternative, data-driven approach to partitioning that directly targets significant price movements within a single series, generating variable-length segments aligned with market phases.

\subsection{Pattern Recognition and Categorization}

Identifying recurring patterns is fundamental to time series analysis. Our approach leverages pattern recognition in two ways: (1) the adaptive segmentation implicitly identifies patterns ending in significant peaks, and (2) the categorization explicitly uses the fixed-length end-pattern of a preceding segment to group subsequent segments. This pattern-conditioned categorization (Section~\ref{subsec:categorization}), based on simple binary sequences derived from volatility changes (Equation~\ref{eq:binary_pattern}), provides a computationally efficient and interpretable way to define context for the specialized TFT models, differing from shape-based clustering methods that aim to group entire time series \cite{paparrizos2015k} or complex probabilistic approaches for market regime discovery \cite{wang2020regime}.

\subsection{Multimodal Forecasting Approaches}

Another significant research direction involves incorporating information beyond historical prices. Multimodal approaches integrate data from various sources, such as technical indicators, macroeconomic variables, or sentiment analysis derived from news headlines or social media. Studies have shown that sentiment data can significantly influence cryptocurrency prices \cite{angela2020factors, ramos2021multi}. Models like MFB \cite{HAN2025125515} explicitly fuse time-lagged sentiment features with technical indicators using BiLSTM and BiGRU layers. Farimani et al. \cite{farimani2021leveraging, farimani2022investigating,farimani2024adaptive,Farimani2025Financial} explored leveraging latent concepts and sentiment from the news using BERT-based transformers. While our current work focuses primarily on adapting the analysis structure based on price/volatility dynamics, integrating external factors like sentiment within our adaptive framework remains a potential avenue for future enhancement.

\subsection{Positioning the Proposed Work}

Our proposed Adaptive TFT Modeling approach builds upon our previous work \cite{Peik2025Enhancing} and introduces a unique combination: it replaces rigid fixed-length windows with adaptive, variable-length segmentation driven by significant market turning points ($T_h$ threshold). It then uniquely conditions the forecasting process by categorizing the \textit{next} segment based on the observed pattern at the \textit{end} of the \textit{previous} segment. By training specialized TFT models for each resulting category, we aim to leverage TFT's forecasting power within a more dynamically relevant and context-aware framework, specifically tailored for short-term cryptocurrency prediction.

\section{Proposed Approach: Adaptive TFT Modeling}
\label{sec:proposed_approach}

\begin{figure}[t]
\centering
\includegraphics[trim=10 20 10 10, clip,width=0.95\hsize]{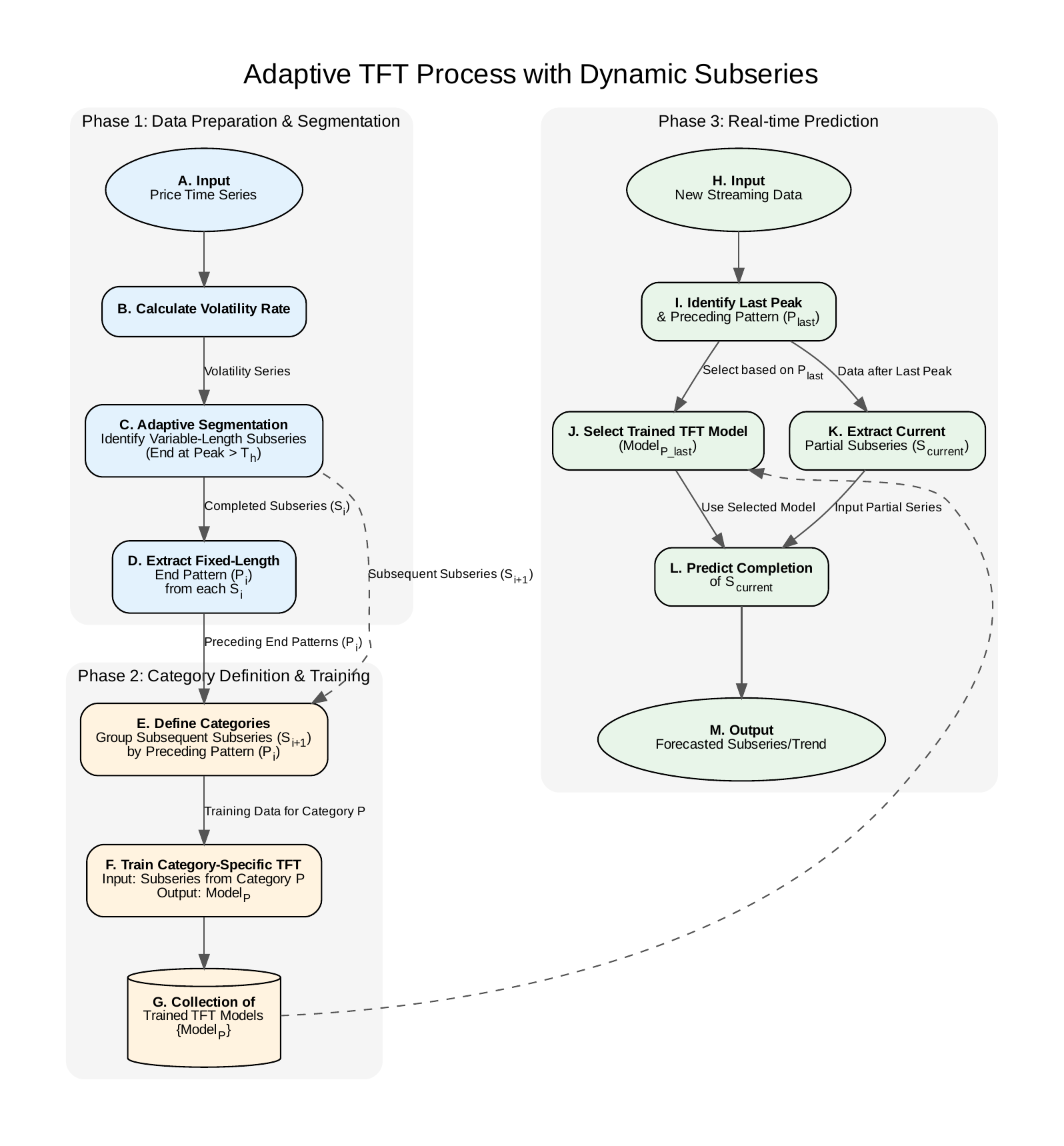}
\caption{Overview of our proposed method.}
\label{fig:adaptive_diagram}
\end{figure}

This section details our proposed methodology for enhancing cryptocurrency price forecasting through an adaptive modeling framework designed to better align with inherent market dynamics. Unlike traditional methods that rely on fixed-length time windows, our approach introduces two key innovations: (1) dynamic segmentation of the time series into variable-length subseries based on significant relative maxima, and (2) pattern-conditioned forecasting where specialized TFT models predict the evolution of a market segment based on the characteristic pattern observed at the end of the preceding segment.By integrating adaptive segmentation and pattern-conditioned forecasting we aim to capture market dynamics more effectively. The overall workflow is depicted in Figure~\ref{fig:adaptive_diagram} including  training (Algorithm~\ref{alg:adaptive_tft_training}) and prediction (Algorithm~\ref{alg:adaptive_tft_prediction}) phases, as elaborated below.

\subsection{Training Phase: Learning from Adaptive Segments and Preceding Patterns}
\label{subsec:training_phase}

The primary goal of the training phase is to construct a library of specialized TFT models $\{Model_P\}$, where each model is an expert at predicting typical market behavior that unfolds after a specific, recurring end pattern $P$ has been observed. The process, described in Algorithm~\ref{alg:adaptive_tft_training}, begins after initial data preparation steps, such as converting raw prices to volatility rates (detailed in Section~\ref{sec:preprocessing}).

First, the historical volatility time series $V$ undergoes \textit{Adaptive Segmentation} (Block~C in Figure~\ref{fig:adaptive_diagram}; line~2 in Algorithm~\ref{alg:adaptive_tft_training}). This core procedure partitions the series $V$ into a sequence of variable-length subseries $\{S_1, S_2, \dots, S_k\}$. The total number of segments, $k$, is determined dynamically by the \texttt{AdaptiveSegment()} function, and its value is data-dependent, varying with the chosen threshold $T_h$. The number of models corresponds to the number of unique end-patterns $P$ found among the $k-1$ completed subseries.

Each subseries $S_i$ terminates at a relative maximum (peak) $t_{peak,i}$. A peak is considered valid only if the magnitude of the increase in price from the preceding relative minimum to this peak exceeds a predefined threshold $T_h$. For example, if $T_h$ is set to 1.5\%, a peak at \$2,050 would only be considered a valid segment endpoint if the preceding trough was below approximately \$2,020, signifying a rise of over 1.5\%. This condition ensures that segmentation points correspond to significant market turning points, potentially filtering out minor noise and aligning the subseries boundaries with meaningful market phases.

\begin{algorithm}[t]
\caption{Adaptive TFT Model Training with Dynamic Subseries}
\label{alg:adaptive_tft_training}
\begin{algorithmic}[1]

\Require Time series $T$, Relative maximum threshold $T_h$, End pattern length $p_{len}$ 

\Ensure Collection of trained TFT models $\{Model_P\}$ for each end pattern $P$

\State $V \leftarrow \text{CalculateVolatilityRate}(T)$ 
\State $\{S_1, S_2, \dots, S_k\} \leftarrow \text{AdaptiveSegment}(V, T_h)$ 
\State Initialize $CategoryData \leftarrow \text{empty dictionary}$ 
\For{$i = 1$ to $k-1$} 
    \State $P_i \leftarrow \text{ExtractEndPattern}(S_i, p_{len})$ 
    \If{$P_i$ not in $CategoryData$}
        \State $CategoryData[P_i] \leftarrow \text{empty list}$ 
    \EndIf
    \State Append $S_{i+1}$ to $CategoryData[P_i]$ 
\EndFor
\State Initialize $TrainedModels \leftarrow \text{empty dictionary}$
\ForEach{Pattern $P$ in $CategoryData.keys()$}
    \State $Data_P \leftarrow CategoryData[P]$
    \State $Model_P \leftarrow \text{TrainTFTModel}(Data_P)$ 
    \State $TrainedModels[P] \leftarrow Model_P$ 
\EndFor

\State \Return $TrainedModels$

\end{algorithmic}
\end{algorithm}

Following segmentation, for each completed subseries $S_i$ (where $i$ ranges from 1 to $k-1$), the \textit{End Pattern Extraction} step (Block~D; line~5 in Algorithm~\ref{alg:adaptive_tft_training}) isolates the final fixed-length sequence of observations, $P_i$, with length $p_{len}$ (e.g. the last 5 time steps). This pattern $P_i$ serves as a signature characterizing the market's state immediately preceding the $i$-th peak.

The subsequent \textit{Categorization} logic (conceptualized in Block~E; lines~6-10 in Algorithm~\ref{alg:adaptive_tft_training}) is crucial: it defines the training data for each specialized model. Specifically, the subseries $S_{i+1}$ (which starts immediately after $S_i$ ends) is assigned to the category associated with the pattern $P_i$. A dictionary, $CategoryData$, is used to collect all subseries that follow the same preceding end-pattern. For instance, all subseries $S_{j+1}$ whose preceding subseries $S_j$ ended with pattern $P$ are grouped into $CategoryData[P]$. This grouping effectively organizes the data based on the consequences of observing specific preceding market conditions.

Finally, for each unique end-pattern $P$ identified, a dedicated \textit{TFT Model is Trained} (Block~F; Algorithm~\ref{alg:adaptive_tft_training}, Lines~13-15). The model $Model_P$ is trained exclusively on the set of variable-length subseries collected in $CategoryData[P]$. This model learns the characteristic temporal dynamics and evolutionary path typical of market segments that begin after pattern $P$ has occurred. The technical details regarding how the TFT architecture handles the variable lengths of these input subseries will be discussed in Section~\ref{subsec:tft_construction}. The culmination of the training phase is the collection of specialized TFT models $\{Model_P\}$ (Block~G), each tuned to predict the behavior conditioned on a specific prior market pattern.

\subsection{Prediction Phase: Forecasting with Adaptive Models}
\label{subsec:prediction_phase}

Once the library of specialized TFT models $\{Model_P\}$ is available, the prediction phase (detailed in Algorithm~\ref{alg:adaptive_tft_prediction} and illustrated in Phase~3 of Figure~\ref{fig:adaptive_diagram}) operates on new, incoming data streams in real-time or near real-time.

\begin{algorithm}[t]
\caption{Adaptive TFT Prediction with Dynamic Subseries}
\label{alg:adaptive_tft_prediction}
\begin{algorithmic}[1] 

\Require New volatility data points $V_{new}$, Historical volatility series $V_{hist}$, Trained TFT models $\{Model_P\}$, Relative maximum threshold $T_h$, End pattern length $p_{len}$
\Ensure Predicted continuation of the current subseries $Prediction$

\State $V_{hist} \leftarrow \text{Append}(V_{hist}, V_{new})$ 

\State $t_{peak} \leftarrow \text{FindLastPeakIndex}(V_{hist}, T_h)$ 

\If{$t_{peak}$ is Null} 
    \State \Return Null 
\EndIf

\State $S_{last} \leftarrow V_{hist}[\dots : t_{peak}]$ 
\State $P_{last} \leftarrow \text{ExtractEndPattern}(S_{last}, p_{len})$ 

\If{$P_{last}$ not in $\{Model_P\}$.keys()} 
    \State \Return Null 
\EndIf
\State $SelectedModel \leftarrow \{Model_P\}[P_{last}]$ 

\State $S_{current\_partial} \leftarrow V_{hist}[t_{peak}+1 : \text{end}]$ 

\If{$\text{Length}(S_{current\_partial}) == 0$} 
    \State \Return Null 
\EndIf

\State $Prediction \leftarrow SelectedModel.\text{predict}(S_{current\_partial})$ 

\State \Return $Prediction$

\end{algorithmic}
\end{algorithm}

As new volatility data points $V_{new}$ arrive (Block~H), they are appended to the historical series $V_{hist}$ (Algorithm~\ref{alg:adaptive_tft_prediction}, Line~1). The system then analyzes the updated $V_{hist}$ to identify the last completed subseries ($S_{last}$) by finding the index of the most recent relative maximum peak ($t_{peak}$) that satisfies the threshold condition $T_h$ (Block~I; Algorithm~\ref{alg:adaptive_tft_prediction}, Line~2).

The end pattern $P_{last}$ is extracted from the final $p_{len}$ steps of this identified $S_{last}$ (line~5 in Algorithm~\ref{alg:adaptive_tft_prediction}). This pattern dictates which specialized model is best suited for the current context. Based on $P_{last}$, the corresponding pre-trained TFT model is selected from the collection, denoted as $SelectedModel = \{Model_P\}[P_{last}]$ (Block~J; Line~8 in Algorithm~\ref{alg:adaptive_tft_prediction}). This ensures that the prediction is made by the model trained on historical sequences that followed the same preceding pattern.

Simultaneously, the system extracts the current partial pubseries $S_{current\_partial}$, which comprises all the data points observed in $V_{hist}$ after the last identified peak $t_{peak}$ (Block~K; Algorithm~\ref{alg:adaptive_tft_prediction}, Line~10). This partial sequence represents the beginning of the currently unfolding market segment.

This initial part of the current subseries, $S_{current\_partial}$, is then fed as input to the $SelectedModel$ (Block~L; Algorithm~\ref{alg:adaptive_tft_prediction}, Line~13). The model leverages its specialized knowledge about the typical evolution following pattern $P_{last}$ to predict the continuation/completion of $S_{current\_partial}$. The prediction typically forecasts the sequence until the next likely relative maximum or for a predefined horizon.

The final output (Block~M) is the predicted trajectory or values for the remainder of the current market segment. This adaptive selection and prediction mechanism allows the system to dynamically choose the most relevant predictive model based on the most recently completed market phase, leading to potentially more accurate and context-aware forecasts.

\subsection{Data Preprocessing}
\label{sec:preprocessing}

The dataset utilized in this study consists of transaction data for the ETH-USDT pair obtained from the Binance exchange, aggregated into 10-minute timeframes. This dataset spans the period from December 2021 to January 2025 (adjust dates as needed) and includes Open, High, Low, Close (OHLC) prices and Volume for each interval. Similar to our previous work \cite{Peik2025Enhancing}, the core feature driving our analysis and modeling is derived from the closing price.

To focus on relative price movements and mitigate issues related to price scale variations over time, we transform the raw closing price series into a \textit{Volatility Rate} series. The volatility rate $V_t$ at timeframe $t$ is calculated as the percentage change from the closing price of the previous timeframe $P_{t-1}$ to the current closing price $P_t$:
\begin{equation}
    V_t = \left( \frac{P_t}{P_{t-1}} - 1 \right) \times 100
    \label{eq:volatility_rate}
\end{equation}
For instance, if a 10-minute closing price sequence is $[3000, 3015, 2995]$, the corresponding volatility rate sequence would be $[0.5, -0.66]$. This transformation yields a time series representing the rate of price change, which serves as the primary input for our adaptive segmentation and subsequent modeling stages.

While other features like volume or technical indicators (e.g., MACD) can often provide valuable information, preliminary analysis (consistent with findings in \cite{Peik2025Enhancing}) suggested that for the specific task of high-frequency pattern categorization and prediction based on our adaptive segmentation, the volatility rate itself captures the most salient dynamics. Therefore, to maintain model simplicity and focus on the core methodology, we primarily utilize the \textit{close price volatility rate} time series for both segmentation and as the target variable for the TFT models.

Although not used as primary segmentation features, time-based covariates, such as the relative position of a timestep within its dynamically determined subseries, can be incorporated as auxiliary inputs to the TFT models. These covariates can help the TFT architecture better understand the temporal context within the variable-length segments (details on TFT inputs are provided in Section~\ref{subsec:tft_construction}).

\subsection{Adaptive Subseries Segmentation}
\label{subsec:adaptive_segmentation}

Traditional fixed-length windowing for time series analysis can arbitrarily cut through meaningful market phases or fails to capture the full duration of a specific pattern. To address this, we introduce an \textit{adaptive segmentation} technique that partitions the volatility rate time series $V = \{v_1, v_2, \dots, v_T\}$ into a sequence of \textit{variable-length} subseries $\{S_1, S_2, \dots, S_k\}$, where each boundary of the subseries aligns with significant market turning points.

The segmentation process relies on identifying specific \textit{relative maxima} (peaks) in the volatility rate series $V$ that signify the culmination of a notable upward price movement. To define these significant peaks, we first need to identify relative minima (troughs). Let $t_{min,last}$ be the index of the last identified relative minimum in the series $V$. A point $v_t$ is considered a potential relative maximum if it is greater than its immediate neighbors (e.g., $v_t > v_{t-1}$ and $v_t > v_{t+1}$, potentially using a slightly wider neighborhood or smoothing for robustness).

However, not every potential relative maximum marks the end of a subseries. We introduce a significance criterion based on the magnitude of the price increase (accumulated volatility) since the last trough. A potential relative maximum at index $t_{peak}$ is confirmed as a valid segment endpoint, marking the end of the current subseries $S_i$ and the beginning of the search for the next minimum, only if the rise from the last identified minimum $v_{t_{min,last}}$ to this peak $v_{t_{peak}}$ exceeds a predefined threshold $T_h$.

The rise, denoted as $Rise(t_{min,last}, t_{peak})$, can be defined as the difference between the peak value and the preceding minimum value on the original price series $P$ (or approximated using the cumulative sum of volatility rates, though direct price comparison is often clearer):
\begin{equation}
    Rise(t_{min,last}, t_{peak}) = \frac{P_{t_{peak}} - P_{t_{min,last}}}{P_{t_{min,last}}} \times 100
    \label{eq:rise_calculation}
\end{equation}
Alternatively, if working purely with volatility rates $V$, one might approximate the condition, although Equation \ref{eq:rise_calculation} using the underlying prices $P$ is more direct for measuring the percentage rise.

The condition for confirming a peak at $t_{peak}$ as a subseries endpoint is then:
\begin{equation}
    Rise(t_{min,last}, t_{peak}) \ge T_h
    \label{eq:threshold_condition}
\end{equation}
where $T_h$ is a positive percentage threshold such as 1\% or 2\%.

The segmentation algorithm proceeds iteratively through the volatility series $V$:
\begin{enumerate}
    \item Initialize the start of the first subseries $S_1$ at $t=1$. Identify the first relative minimum $t_{min,last}$.
    \item Scan forward from $t_{min,last}$ to find the next potential relative maximum at index $t_{potential\_peak}$.
    \item Calculate $Rise(t_{min,last}, t_{potential\_peak})$ using Equation~\ref{eq:rise_calculation}.
    \item Check the threshold condition (Equation~\ref{eq:threshold_condition}):
        \begin{itemize}
            \item If $Rise \ge T_h$: Confirm $t_{potential\_peak}$ as the end index of the current subseries $S_i$. The subseries is $S_i = \{v_{t_{start,i}}, \dots, v_{t_{potential\_peak}}\}$. Set $t_{start,i+1} = t_{potential\_peak} + 1$. Find the next relative minimum after $t_{potential\_peak}$ and update $t_{min,last}$. Repeat from step 2 for the next subseries $S_{i+1}$.
            \item If $Rise < T_h$: This peak is not significant enough. Disregard $t_{potential\_peak}$ as an endpoint. Continue scanning from $t_{potential\_peak}+1$ to find the next potential relative maximum (step 2), keeping the same $t_{min,last}$. If a new, lower relative minimum is found before a valid peak, update $t_{min,last}$.
        \end{itemize}
    \item Continue until the end of the time series $V$ is reached.
\end{enumerate}

This process yields a sequence of subseries $S_1, S_2, \dots, S_k$, where each $S_i$ represents a market phase ending in a significant upward thrust defined by the threshold $T_h$. The lengths of these subseries are inherently variable, adapting to the natural rhythm of the market's advances and consolidations. The threshold $T_h$ acts as a sensitivity parameter; a higher $T_h$ results in longer, fewer subseries capturing only major movements, while a lower $T_h$ leads to shorter, more numerous subseries sensitive to smaller fluctuations. The choice of $T_h$ is crucial and can be optimized based on the specific cryptocurrency, timeframe, and modeling goals.

\subsection{Pattern-Based Subseries Categorization}
\label{subsec:categorization}

Following the adaptive segmentation process described in Section~\ref{subsec:adaptive_segmentation}, which yields variable-length subseries $\{S_1, S_2, \dots, S_k\}$ ending at significant relative maxima, the next crucial step is to categorize these market phases to enable specialized model training. Unlike traditional clustering methods such as $k$-means that rely on geometric distance, our approach employs a direct categorization strategy based on market dynamics. Such clustering techniques can erroneously group subseries that are behaviorally opposite but geometrically close; for example, a subseries representing a slow price ascent could be clustered with one showing a slow descent. To avoid this pitfall, our direct categorization strategy instead groups subseries based on the market dynamics immediately preceding their start, which ensures that each category is contextually consistent.

The core idea is that the way a market phase (subseries $S_i$) concludes often influences the characteristics of the phase that follows ($S_{i+1}$). Therefore, we categorize the subseries $S_{i+1}$ based on the end pattern $P_i$ extracted from the final fixed number of steps of the preceding subseries $S_i$.

Specifically, for each completed subseries $S_i$ (where $i < k$), we extract its end pattern $P_i$ consisting of the sequence of volatility rates in its last $p_{len}$ time steps:
\begin{equation}
    P_i = \{ v_{t_{peak,i}-p_{len}+1}, v_{t_{peak,i}-p_{len}+2}, \dots, v_{t_{peak,i}} \}
\end{equation}
where $t_{peak,i}$ is the index of the relative maximum ending subseries $S_i$.

To create discrete categories, this pattern $P_i$ is transformed into a binary representation based on the direction of change between consecutive steps. Let $b_t$ represent the binary state at time $t$ within the pattern ($p_{len}-1$ states in total):
\begin{equation}
    b_t = 
    \begin{cases} 
      1 & \text{if } v_t \ge v_{t-1} \quad (\text{Increase or No Change}) \\
      0 & \text{if } v_t < v_{t-1} \quad (\text{Decrease}) 
    \end{cases}
    \label{eq:binary_pattern}
\end{equation}
for $t$ ranging from $t_{peak,i}-p_{len}+2$ to $t_{peak,i}$. This results in a binary sequence of length $p_{len}-1$ representing the end pattern $P_i$. For example, if $p_{len}=5$, the pattern consists of 4 binary digits.

All subseries $S_{i+1}$ that follow a preceding subseries $S_i$ ending with the $same$ binary pattern $P$ are grouped into the same category, denoted as $Category P$. The total number of distinct categories is therefore $2^{p_{len}-1}$. If $p_{len}=5$, this results in $2^4 = 16$ distinct categories.

This categorization method offers several advantages over traditional clustering for our purpose:
\begin{itemize}
    \item \textbf{Behavioral Relevance:} It groups subsequent market phases based on the specific, recent historical pattern that preceded them, which is hypothesized to be more relevant for predicting the immediate future than overall geometric similarity.
    \item \textbf{Computational Efficiency:} It avoids the computationally intensive process of clustering high-dimensional, variable-length time series data. Categorization is achieved through direct pattern extraction and mapping.
    \item \textbf{Interpretability:} Each category has a clear definition based on a specific sequence of recent up/down movements.
\end{itemize}

As outlined in Algorithm~\ref{alg:adaptive_tft_training} (lines 4-10), this categorization process is used during the training phase to organize the subseries $S_{i+1}$ into distinct datasets ($CategoryData[P]$), each corresponding to a unique preceding end pattern $P$. These categorized datasets then form the basis for training the specialized TFT models described in the following section.

\subsection{Temporal Fusion Transformer (TFT) Architecture}
\label{subsec:tft_construction}

In order to predict the evolution of the variable-length subseries within each identified category, we employ the TFT model \cite{lim2021temporal}, a deep learning architecture specifically designed for interpretable multi-horizon time series forecasting capable of handling diverse input types and capturing complex temporal dependencies across different scales. It builds on the success of transformer models \cite{vaswani2017attention} in sequence processing by incorporating mechanisms tailored for time series data.

The TFT architecture integrates several specialized components:

\begin{enumerate}
    \item \textbf{Input Processing and Variable Selection:} TFT is designed to handle three types of inputs:
        \begin{itemize}
            \item \textit{Static Covariates:} Features that do not change over time (e.g., category ID $P$ in our case, although not explicitly used as input here, it determines which model is used). These are typically processed using entity embeddings.
            \item \textit{Time-Varying Known Inputs:} Features whose future values are known (e.g., day of the week, time of day, relative position within a known forecast horizon).
            \item \textit{Time-Varying Observed Inputs:} The primary time series values (e.g., our volatility rates $V$) and any other covariates known only up to the present.
        \end{itemize}
    A key component is the {Variable Selection Network (VSN)}, applied to static, past, and future inputs. VSNs learn the relevance of each input variable, assigning weights to prioritize the most salient features for the specific forecasting task and time step. This is often implemented using Gated Residual Networks (GRNs) acting on transformed inputs, followed by a softmax layer to produce weights $\mathbf{w}$:
    \begin{equation}
        {VSN}(\mathbf{x}) = {Softmax}({GRN}(\mathbf{x})) \odot {GRN}(\mathbf{x})
    \end{equation}
    where $\mathbf{x}$ represents the input features and $\odot$ denotes element-wise multiplication. This allows the model to dynamically emphasize or de-emphasize different inputs.

    \item \textbf{Sequence-to-Sequence Processing with LSTMs:} The core temporal dynamics are captured using an LSTM-based encoder-decoder layer. This layer processes the time-varying inputs (after variable selection and transformation) to learn sequential patterns and dependencies. Unlike standard Transformers that rely solely on attention, TFT retains LSTMs for robust handling of long-range temporal order.

    \item \textbf{Static Enrichment:} Information from static covariates is integrated into the temporal processing through context vectors. These vectors are generated using GRNs applied to the static features and are used to condition the LSTM outputs, allowing static information to influence the temporal feature extraction at each time step.

    \item \textbf{Interpretable Multi-Head Attention:} Following the LSTM layer, TFT employs a modified multi-head self-attention mechanism, similar to the original Transformer \cite{vaswani2017attention}. This allows the model to assess the importance of past time steps when making predictions for the future. For a set of input representations $H = [\mathbf{h}_1, \dots, \mathbf{h}_N]$, the attention output is calculated as:
    \begin{equation}
        {Attention}(Q, K, V) = {softmax}\left(\frac{QK^T}{\sqrt{d_k}}\right)V
    \end{equation}
    where $Q = HW_Q$, $K = HW_K$, $V = HW_V$ are the queries, keys, and values obtained via linear projections ($W_Q, W_K, W_V$), and $d_k$ is the dimension of the keys. The attention weights from $\textit{softmax}(\dots)$ provide insights into which past time steps were most influential, contributing to the model's interpretability. The "multi-head" aspect allows attending to different temporal patterns simultaneously.

    \item \textbf{Gated Residual Network (GRN):} GRNs are fundamental building blocks used throughout the TFT architecture (in VSNs, static enrichment, and attention layers). They provide flexible non-linear processing while incorporating skip connections for stable training. A GRN takes a primary input $\mathbf{a}$ and an optional context vector $\mathbf{c}$:
    \begin{align}
        {GRN}(\mathbf{a}, \mathbf{c}) &= {LayerNorm}(\mathbf{a} + {GatedLinearUnit}(\eta_1)) \\
        \eta_1 &= W_1 \eta_2 + b_1 \\
        \eta_2 &= {ELU}(W_2 \mathbf{a} + W_3 \mathbf{c} + b_2)
    \end{align}
    where $W_1, W_2, W_3, b_1, b_2$ are weights and biases, ELU is the activation function, and \textit{GatedLinearUnit} (GLU) applies a sigmoid gate: ${GLU}(x) = \sigma(W_g x + b_g) \odot (W_v x + b_v)$.

    \item \textbf{Prediction Head:} The output from the attention layer is passed through another GRN and finally through linear layers to produce the forecasts. TFT is often configured to output quantile forecasts, providing prediction intervals rather than just point estimates.
\end{enumerate}

\textbf{Handling Variable-Length Subseries:} A critical consideration for our methodology is that the subseries $S_{i+1}$ used to train each $Model_P$ have variable lengths, determined by adaptive segmentation. Standard TFT implementations often expect fixed-length inputs. To accommodate this, techniques such as \textbf{padding and masking} are typically employed:
\begin{itemize}
    \item \textbf{Padding:} All input sequences (both historical inputs and known future inputs, if any) within a batch are padded to the length of the longest sequence in that batch using a special value such as zero.
    \item \textbf{Masking:} Corresponding masks are generated to indicate which elements are real data and which are padding. These masks are used within the LSTM layers and, crucially, within the self-attention mechanism (specifically in the softmax calculation) to ensure that the model does not attend to or process the padded values.
\end{itemize}
Implementations like the TFT model available in some libraries \cite{darts2022} often provide mechanisms to handle masking implicitly when fed sequences of varying lengths, simplifying the process for the user. This allows our specialized $Model_P$ to learn effectively from the variable-length subseries characteristic of Category $P$.

In our framework, each $Model_P$ is trained using the corresponding categorized dataset $Data_P$ (Algorithm~\ref{alg:adaptive_tft_training}, Line 14). The goal is for $Model_P$ to learn the specific temporal patterns associated with subseries that follow the end pattern $P$, leveraging the architectural strengths of TFT described above to make accurate predictions for the continuation of the new partially observed subseries identified as belonging to Category $P$ during the prediction phase.

\section{Experiments and Results}
\label{sec:experiments}

In this section, we evaluate the performance of our proposed Adaptive TFT Modeling approach for short-term cryptocurrency price forecasting. We compare its effectiveness against several baseline methods using a real-world dataset.

\subsection{Dataset}
\label{subsec:dataset}

The empirical basis for this investigation is a proprietary, high-resolution dataset of 1-minute ETH-USDT trading intervals, collected from the Binance exchange for the period of December 27, 2021, to November 22, 2024. This dataset is one of the most extensive of its kind at this granularity and we made it publicly available at \cite{githubGitHubArashitc2Binance1minutecandles}. The high-frequency nature of the data permits aggregation to any analytical timeframe; for this study, the 1-minute intervals were resampled into 10-minute periods, from which Open, High, Low, Close (OHLC) prices and Volume were extracted. Following the methodology detailed in Section~\ref{sec:preprocessing}, the closing prices were transformed into a volatility rate series (Equation~\ref{eq:volatility_rate}) to serve as the primary input for our models. A chronological split was employed, designating data up to November 14, 2024, for model training, with the final week from November 15 to November 22, 2024, reserved for out-of-sample validation. Notably, this validation period was characterized by abnormal upward volatility, providing a challenging scenario to evaluate the model's robustness against market conditions where traditional models often underperform.

\subsection{Experimental Setup}
\label{subsec:exp_setup}

\subsubsection{Implementation Details}
Our proposed adaptive TFT model and the baseline models were implemented in Python, leveraging the Darts library \cite{darts2022} for its comprehensive suite of time series forecasting models, including TFT. To ensure the reproducibility of our findings and to facilitate further research, the complete source code for our adaptive methodology and all reported experiments are publicly available \cite{githubSourceCode}. The key hyperparameters for our adaptive method, such as the relative maximum threshold ($T_h$) and the end-pattern length ($p_{len}$), were determined through empirical validation and are detailed within the repository.

\subsubsection{Baseline Models}
To assess the effectiveness of our adaptive approach, we compare its performance against the following baseline models:
\begin{itemize}
    \item \textbf{Standard LSTM:} A standard Long Short-Term Memory network trained directly on the volatility rate series using a fixed lookback window, implemented based on \cite{song2020time}.
    \item \textbf{Standard TFT:} The Temporal Fusion Transformer model \cite{lim2021temporal} trained directly on the volatility rate series with fixed-length input windows, without adaptive segmentation or categorization.
    \item \textbf{Fixed-Length Categorized TFT (FL-Cat-TFT):} Our previous approach \cite{Peik2025Enhancing}, which uses fixed-length subseries, categorizes them based on internal patterns, and employs specialized TFT models with a selector mechanism. The original implementation for this baseline model is publicly available for reproducibility \cite{githubGitHubArashitc2sourcecode}.
\end{itemize}

The hyperparameters for the baseline models were tuned to ensure a fair comparison.

\subsubsection{Evaluation Metrics}
We evaluate the models based on two primary aspects:
\begin{itemize}
    \item \textbf{Predictive Accuracy:} We measure the model's ability to correctly predict the direction (up or down) of the volatility rate in the next 10-minute timeframe. We report:
        \begin{itemize}
            \item \textbf{Accuracy:} The overall percentage of correct directional predictions.
            \item \textbf{Precision (for upward trend):} The proportion of predicted upward movements that were actually upward movements. This is crucial for spot trading strategies to avoid buying before a price drop.
        \end{itemize}
    \item \textbf{Simulated Trading Profitability:} We simulate a simple spot trading strategy on the ETH-USDT pair during the test period. Starting with an initial capital (e.g. 100 USDT), a buy signal is generated if the model predicts an upward trend for the next timeframe, and a sell signal is generated if it predicts a downward trend (or holds if already sold). We assume zero transaction fees for simplicity. The final asset value at the end of the test period reflects the practical utility of the model.
\end{itemize}

\subsection{Performance Evaluation}
\label{subsec:performance}

The performance of our proposed adaptive TFT model compared to the baseline methods on the ETH-USDT test set is summarized in Table~\ref{tab:accuracy_precision} for predictive accuracy and Table~\ref{tab:profitability} for simulated trading profitability.

\begin{table}[t]
\centering
\caption{Model Performance Comparison on the ETH-USDT Test Period.}
\label{tab:accuracy_precision}
\begin{tabular*}{\columnwidth}{@{\extracolsep{\fill}}lcccc@{}}
\toprule
Model                        & Accuracy & Precision & Recall & Specificity \\ 
                             & (\%)     & (\%)      & (\%)   & (\%)        \\ \midrule
Standard LSTM                & 49.15    & 49.90     & 49.80  & 49.70       \\
Standard TFT                 & 47.75    & 48.90     & 48.61  & 48.90       \\
FL-Cat-TFT (\cite{Peik2025Enhancing})   & 50.32    & 50.11     & 58.72  & 52.79       \\
\textbf{Adaptive TFT} & \textbf{51.36} & \textbf{51.11} & \textbf{92.31} & \textbf{19.28} \\ \bottomrule
\end{tabular*}
\end{table}

\begin{table}[t]
\centering
\caption{Simulated Trading Profitability Comparison on ETH-USDT (Test Period, Initial Capital: 100 USDT).}
\label{tab:profitability}
\begin{tabular}{@{}lc@{}}
\toprule
Strategy / Model                     & Final Asset Value (USDT) \\ \midrule
Buy and Hold                         & 108.32                \\
Standard LSTM Trading                & 112.43                \\
Standard TFT Trading                 & 102.90                \\
FL-Cat-TFT Trading (\cite{Peik2025Enhancing})   & 114.07                \\
\textbf{Adaptive TFT Trading} & \textbf{117.22}   \\ \bottomrule
\end{tabular}
\end{table}

As shown in Table~\ref{tab:accuracy_precision}, the proposed Adaptive TFT model achieves an accuracy of 51.36\% and a precision of 51.11\%, outperforming all baseline models. This suggests that the adaptive segmentation and pattern-conditioned forecasting lead to more reliable directional predictions. The results do not show high values for accuracy and precision, and the improvement does not seem to be considerable. However, note that we restricted our time frame to 10 minutes, which is small compared to
hourly/daily time frames. Cryptocurrency markets are chaotic in small time frames and
due to the large number of transactions any small improvement in accuracy and precision
 can have a great impact on long-term profits. Our strategy takes advantage of short-term market movements and captures quick profits in volatile markets. We  explain the importance of the results in Section \ref{sec:discussion}.

Furthermore, the simulated trading results in Table~\ref{tab:profitability} indicate the practical benefit of the improved accuracy. Our Adaptive TFT model resulted in a final asset value of 117.22 USDT, demonstrating higher profitability compared to both the passive Buy and Hold strategy and the trading strategies based on the baseline predictive models. These results support our hypothesis that adapting the model structure to market dynamics through variable-length segmentation and pattern conditioning enhances forecasting performance for cryptocurrency trading.

\section{Discussion}
\label{sec:discussion}

The experimental results presented in Section~\ref{sec:experiments} provide compelling evidence for the effectiveness of our proposed adaptive TFT modeling approach. As shown in Table~\ref{tab:profitability}, the method achieved the highest simulated trading profitability (117.22 USDT), significantly outperforming the baseline models. A closer look at the results in Table~\ref{tab:accuracy_precision} reveals that this superior financial performance is not due to an increase in overall accuracy (51. 36\%), but because of a distinct and aggressive predictive strategy. Our adaptive model exhibits an exceptionally high recall (92.31\%), indicating the successful identification of nearly all true upward price movements. This bullish conviction, however, comes at the cost of a low specificity (19.28\%), suggesting that the model frequently classified minor downward or flat periods as continuations of an upward trend. This specific trade-off is shown to be a highly effective strategy within the volatile, upward-trending market of our test period. This section discusses the implications of these findings, compares our approach with related work, and acknowledges the current limitations.

\subsection{Interpretation of Performance Gains}

The enhanced performance of the Adaptive TFT model can be attributed primarily to the synergy between its two core components: \textit{adaptive segmentation} and \textit{pattern-conditioned forecasting}.

Firstly, the \textit{adaptive segmentation} based on thresholded relative maxima ($T_h$) allows the model to partition the time series into variable-length segments that likely correspond more closely to natural market phases than arbitrary fixed-length windows. By defining segment boundaries based on significant upward movements, the method potentially captures complete trends or cycles, providing more coherent inputs for pattern analysis and forecasting. This contrasts with fixed windows, which might cut through such phases, mixing different dynamics within a single input sequence.

Secondly, the \textit{pattern-conditioned categorization} strategy introduces a novel way to specialize the predictive models. By categorizing a subseries $S_{i+1}$ based on the specific pattern $P_i$ observed at the end of the preceding subseries $S_i$, we train each TFT model $Model_P$ to become an expert on the typical market evolution that follows the pattern $P$. This differs significantly from our previous work (FL-Cat-TFT) where categorization was based on the internal structure of the fixed-length subseries itself. The current approach hypothesizes that the immediate past pattern is a strong conditioner for near-future behavior, allowing $Model_P$ to make more context-aware predictions for the initial steps of the subsequent subseries. The results suggest this conditioning leads to better performance than both generic models (Standard LSTM/TFT) and the previous fixed-length categorization approach.

A noteworthy observation from Table~\ref{tab:accuracy_precision} is the lower specificity score of our proposed adaptive TFT model compared to the baselines. Specificity, or the true negative rate, measures a model's ability to correctly identify non-upward trends, thereby avoiding erroneous buy signals. In typical market conditions, a high specificity is crucial for robust risk management. However, the model's lower specificity in this case must be interpreted within the context of the validation period, which, as noted in Section~\ref{subsec:dataset}, was characterized by abnormal upward volatility. Our analysis indicates that the adaptive TFT model successfully captured the dominant bullish macro-trend. Consequently, it treated many minor, short-lived downward fluctuations not as trend reversals, but as noise within the larger upward trajectory. While this behavior inherently reduces the true negative rate and thus lowers the specificity score, it proved to be a highly effective strategy in this specific market regime. By maintaining its conviction in the primary trend and not overreacting to insignificant dips, the model avoided premature exits from profitable positions, which directly contributed to its superior trading performance shown in Table~\ref{tab:profitability}. This outcome highlights a trade-off made by the model, which sacrifices some accuracy on minor negative movements to maximize gains from the correctly identified dominant market trend, demonstrating a strong adaptive capability.

\subsection{Significance in High-Frequency Context}

While the absolute percentage gains in accuracy or profitability might seem modest, their significance is amplified in the context of high-frequency (10-minute intervals) cryptocurrency trading. These markets exhibit high noise and near-chaotic behavior, making consistent prediction notoriously difficult \cite{corbet2019cryptocurrencies}. Even small, statistically consistent improvements in directional accuracy or precision, when applied to numerous trading opportunities, can compound into substantial financial gains. The ability of our adaptive model to outperform established baselines like TFT and LSTM, as well as our prior categorized approach, indicates a tangible advancement in capturing relevant short-term dynamics.

\subsection{Role of Parameters ($T_h$ and $p_{len}$)}

The performance of adaptive segmentation is dependent on the choice of the relative maximum threshold $T_h$. This parameter controls the sensitivity of the segmentation: a lower $T_h$ identifies more frequent, smaller peaks, leading to shorter, more numerous subseries, while a higher $T_h$ focuses only on major peaks, resulting in longer, fewer subseries. Finding an optimal $T_h$ involves balancing the need to capture sufficient market phases against the risk of segmenting based on noise. Similarly, the end-pattern length $p_{len}$ determines the specificity of the conditioning patterns and the number of resulting categories ($2^{p_{len}-1}$). A larger $p_{len}$ creates more specific contexts, but risks data sparsity for less common patterns, potentially hindering the training of some specialized TFT models. Conversely, a smaller $p_{len}$ ensures more data per category but might provide less discriminative conditioning information. The optimal values for $T_h$ and $p_{len}$ likely depend on the specific asset and timeframe and require careful tuning, representing an important aspect of applying this methodology.

\subsection{Comparison with Related Approaches}

Compared to regime-switching models such as RHINE \cite{xu2024rhine}, which focus on discovering underlying market states often across multiple series using sophisticated representation learning, our approach takes a more direct predictive stance focused on intra-series dynamics. We do not explicitly model regimes but rather use adaptive segmentation to define phases and pattern-conditioning to specialize TFT forecasts for the subsequent phase. While RHINE provides insights into broader market states, our method is tailored towards generating actionable, short-term directional predictions based on recent, localized patterns.

Compared to multimodal approaches such as MFB \cite{HAN2025125515} that integrate external data (e.g. sentiment), our current model relies solely on price-derived volatility. While this simplifies the model, it potentially misses valuable predictive signals from other sources.

\subsection{Limitations and Future Work}
\label{subsec:limitations_future}

Despite promising results, this study has several limitations that suggest avenues for future research, inspired by the challenges identified in this and previous work \cite{Peik2025Enhancing}:

\begin{itemize}
    \item \textbf{Generalizability and Robustness:} The evaluation was conducted primarily on the ETH-USDT pair and a specific timeframe (10-minute intervals). Further testing across a diverse range of cryptocurrencies (with varying volatility, liquidity, market cap) and potentially different timeframes (including longer-term analysis, perhaps drawing on fractal analysis concepts \cite{bhatt2015fractal}) is necessary to confirm the generalizability and robustness of the adaptive approach.
    \item \textbf{Parameter Optimization:} A systematic investigation into the optimal selection of the segmentation threshold $T_h$ and the end-pattern length $p_{len}$ is warranted. Exploring data-driven methods, adaptive parameter values that change with market conditions, or multi-objective optimization could yield significant improvements.
    \item \textbf{Incorporating External Factors and Covariates:} The current model relies solely on volatility rates derived from price. Integrating relevant external data sources as covariates within the TFT models – such as trading volume, order book information, technical indicators (e.g., MACD, RSI), fundamental data related to the specific cryptocurrency, or market sentiment derived from news and social media \cite{HAN2025125515, farimani2022investigating} – could potentially enhance predictive accuracy significantly.
    \item \textbf{Handling Data Sparsity for Categories:} The pattern-conditioned categorization requires sufficient occurrences of each end-pattern $P$ to adequately train the corresponding TFT model $Model_P$. Less frequent patterns might lead to poorly trained or unreliable models. Future work could explore techniques for handling data sparsity, such as grouping similar patterns, using transfer learning between models, or employing data augmentation techniques specific to time series.
    \item \textbf{Advanced Segmentation Criteria:} While thresholded relative maxima provide a data-driven approach, exploring alternative adaptive segmentation criteria could be beneficial. This might include methods based on volatility regime changes, statistical structural break detection, or clustering-based segmentation applied to short windows.
    \item \textbf{Optimizing TFT for Variable Lengths:} While padding and masking are standard techniques for handling variable-length sequences in TFTs \cite{wang2024timexer}, further research into optimizing TFT architectures, attention mechanisms, or positional encodings specifically for the variable-length financial subseries generated by our method could yield improvements.
    \item \textbf{Alternative Phase Definitions:} Our current work defines a market phase as a "sentence" ending with a significant relative maximum. A promising direction for future research is to explore alternative definitions for these phases, such as identifying movements that conclude at significant relative minima (troughs). This would allow the framework to model different types of market narratives, such as capitulation or accumulation periods, potentially capturing a more diverse set of predictive patterns.
    \item \textbf{Trading Strategy Optimization:} The simulated trading was based on simple directional signals. Developing more sophisticated trading strategies that incorporate the magnitude of predicted changes, confidence levels from the TFT (e.g. quantile forecasts), risk management rules, and potentially reinforcement learning could translate the predictive gains into more substantial and robust trading profitability, especially for leveraged trading scenarios.
\end{itemize}
Addressing these limitations will be crucial to developing more robust, generalizable, and practical adaptive forecasting models for the dynamic cryptocurrency market.

\section{Conclusion}
\label{sec:conclusion}

This paper addressed the challenge of short-term cryptocurrency price forecasting by proposing a novel Adaptive Temporal Fusion Transformer (TFT) Modeling approach. We introduced a dynamic segmentation technique based on thresholded relative maxima to partition the volatility time series into variable-length subseries, better aligning with natural market phases compared to traditional fixed-window methods. Furthermore, we presented a pattern-conditioned forecasting strategy where specialized TFT models are trained for categories defined by the end-pattern of the preceding market segment.

Our experimental evaluation on 10-minute ETH-USDT data demonstrated that this adaptive methodology significantly outperforms standard LSTM, standard TFT, and our previous fixed-length categorization approaches in terms of both directional prediction accuracy and simulated trading profitability. The results suggest that dynamically segmenting the time series and conditioning predictions on recent historical patterns allows the TFT models to capture the complex, non-stationary dynamics of the cryptocurrency market more effectively.

By adapting the analysis structure to the market's own rhythm and specializing predictive models based on context, our approach offers a promising direction for developing more robust and responsive forecasting tools for high-frequency cryptocurrency trading. Although more  research is needed to explore parameter optimization, generalizability, and integration of external factors, as outlined in Section~\ref{subsec:limitations_future}, our proposed adaptive TFT modeling framework represents a valuable contribution towards more accurate and practical cryptocurrency price prediction.

\bibliographystyle{unsrt}  
\bibliography{references}  

\end{document}